\documentstyle[12pt,fleqn,epsf]{article}
\renewcommand{\baselinestretch}{1.2}

\def\fnote#1#2{\begingroup\def\thefootnote{#1}\footnote{#2}\endgroup}
\makeatletter
\def\section{\@startsection {section}{1}{\z@}{3.5ex plus 1ex minus
    .2ex}{2.3ex plus .2ex}{\sc }}
\def\subsection{\@startsection{subsection}{2}{\z@}{3.25ex plus 1ex
minus
   .2ex}{1.5ex plus .2ex}{\small \sc }}
\def\appendix{\par\clearpage
  \setcounter{section}{0}
  \setcounter{subsection}{0}
  \@addtoreset{equation}{section}
  \def\@sectname{Appendix~}
  \def\theequation{\thesection.\arabic{equation}}
  \def\thesection{\Alph{section}}}
\makeatother
\makeatletter \@addtoreset{equation}{section} \makeatother
\renewcommand{\theequation}{\thesection.\arabic{equation}}

\def\ap#1#2#3{     {\it Ann. Phys. (NY) }{\bf #1} (19#2) #3}

\def\npb#1#2#3{    {\it Nucl. Phys. }{\bf B #1} (19#2) #3}
\def\plb#1#2#3{    {\it Phys. Lett. }{\bf B #1} (19#2) #3}
\def\prd#1#2#3{    {\it Phys. Rev. }{\bf D #1} (19#2) #3}
\def\pr#1#2#3{    {\it Phys. Rev. }{\bf #1} (19#2) #3}

\def\prl#1#2#3{    {\it Phys. Rev. Lett. }{\bf #1} (19#2) #3}

\def\zpc#1#2#3{    {\it Z. Physik }{\bf C #1} (19#2) #3}

\def\nc#1#2#3{     {\it Nuovo Cim. }{\bf #1} (19#2) #3}

\def\ijmpa#1#2#3{  {\it Int. J. Mod. Phys. }{\bf A #1} (19#2) #3}

\def\eq#1{{eq.~(\ref{#1})}}
\def\eqs#1#2{{eqs.~(\ref{#1})--(\ref{#2})}}

\let\vev\VEV

\def\Re{\mathop{\mbox{Re}}}

\def\etal{{\it et al.}}

\newcommand{\bea}{\begin{eqnarray}}
\newcommand{\beq}{\begin{equation}}
\newcommand{\eea}{\end{eqnarray}}
\newcommand{\eeq}{\end{equation}}

\newcommand{\spav}[1]{\parbox{1mm}{\vspace*{#1}}}

\begin{document}
\begin{titlepage}
\begin{flushright}
{\tt SISSA 102/95/EP}
\end{flushright}
\spav{1cm}
\begin{center}
{\Large\bf The $\Delta I = 1/2$ Selection Rule}\\
\spav{1cm}\\
{\large V. Antonelli, S. Bertolini, 
M. Fabbrichesi and
E.I. Lashin\fnote{\dag}{Permanent address:
Ain Shams University, Faculty of Science, Dept. of Physics, Cairo, Egypt.}}
\spav{1.5cm}\\
{\em  INFN, Sezione di Trieste}\\
{\em and}\\ 
{\em Scuola Internazionale Superiore di Studi Avanzati}\\
{\em via Beirut 4, I-34013 Trieste, Italy.}\\
\spav{1.5cm}\\
{\sc Abstract}
\end{center}
We compute the isospin $I= 0$ and 2 amplitudes for the decay of a kaon
into two pions by estimating the
relevant hadronic matrix elements in the chiral quark model. The results are
parametrized in terms of  the quark and gluon condensates and of the
constituent quark mass $M$. The latter is a parameter characteristic 
of the model, that we restrict by matching the results in the two 
$\gamma_5$-schemes (HV and NDR) of dimensional regularization.
We find that, for values of  these parameters within the current
determinations,  the $\Delta I= 1/2$ selection rule is well reproduced 
by means of the cumulative effects of short-distance NLO
Wilson coefficients, penguin diagrams, non-factorizable soft-gluon 
corrections and meson-loop renormalization. 
 
 \vfill
\spav{.5cm}\\
{\tt SISSA 102/95/EP}\\
{\tt  October 1995 }

\end{titlepage}

\newpage
\setcounter{footnote}{0}
\setcounter{page}{1}

\section{Introduction}

 The $\Delta I = 1/2$ selection rule~\cite{rule} states that, in the weak
 non-leptonic decays of kaons (as well as of hyperons) the amplitude in
 which the change in isospin is 3/2 is very suppressed with respect to
 that in which the change is 1/2.
  
For the decay of a neutral kaon into two
pions, the $CP$-conserving amplitude
with a final $I=0$ state ($\Delta I = 1/2$)
is measured to be~\cite{PDB} 
\beq
\Re \, A_0 ( K^0 \rightarrow 2 \pi) = 3.33 \times 10^{-7} \: 
\mbox{GeV}\, , 
\eeq
and it is approximately 22 times larger than that
with the pions in the $I=2$ state ($\Delta I = 3/2$):
\beq
\Re \, A_2 ( K^0 \rightarrow 2 \pi)  = 1.50 \times 10^{-8}\: 
\mbox{GeV} \, . 
\eeq
Since a naive estimate of the relevant
hadronic matrix elements within the standard model 
leads to amplitudes that are comparable in size, this 
selection rule has been a standing
puzzle, the solution of which has attracted a great deal 
of theoretical work 
over the past 40 years (for a review see, for instance, ref. \cite{1/2}).

Among the most important steps in the progressive understanding of 
this selection rule,
we mention (without any pretense to completeness) the work of 
Wilson~\cite{Wilson}, Gaillard--Lee~\cite{GL} and 
Altarelli--Maiani~\cite{AM} who stressed the importance of 
QCD short-distance
corrections; the contribution 
of Vain\-shtein--Za\-kh\-arov--Shif\-man~\cite{SVZ} who
identified in the 
gluon penguin operators a possible source of enhancement of $A_0$ 
with respect of $A_2$, 
and the subsequent work of various authors~\cite{penguins} who correctly
estimated the size of them.

At the same time, a parallel development took place as 
Cohen--Manhoar~\cite{CM} first
pointed out the relevance for the selection rule of 
meson-loop corrections in a chiral quark model
computation, Bardeen--Buras--Gerard~\cite{BBG} estimated this effect in the
$1/N_c$ approach with 
the meson loops regularized by a cutoff and, more recently,  
Kambor--Missimer--Wyler~\cite{meson} in chiral perturbation theory. 

What is  then the present status of the rule? 

The two main ingredients entering the theoretical estimate of the
selection rule 
are the perturbative determination of the Wilson coefficients
of the relevant operators and  
the (non-perturbative) evaluation of the hadronic matrix elements. 

The short distance part of the analysis
can be obtained by a straigthforward (albeit challenging) application
of renormalization group equation methods. 
In this respect, we have only updated 
existing analyses by including 
next-to-leading order results~\cite{Monaco,Roma} and evaluating
the Wilson coefficients at various hadronic scales for an updated range 
of $\Lambda_{\rm QCD}^{(4)}$, as explained in section 1.  

Most of the uncertainty resides 
in the second ingredient: the evaluation of hadronic matrix
elements. The vacuum saturation approximation (VSA) is not
sufficient---non-factorizable and non-perturbative effects are
essential in a dynamical explanation of the rule---and
a model for QCD at low energies is called for in order
to make a progress in this direction.

In ref.~\cite{HME} (hereafter referred to as I) we have completed
a systematic study of the chiral
quark model ($\chi$QM)~\cite{QM,EdeRT,ENJL} and 
derived the complete
$O(p^2)$ $\Delta S=1$ chiral Lagrangian. The model allows us to
determine
the coefficient of each term of
the chiral lagrangian  as a function of a few input parameters.
All the relevant hadronic matrix elements can accordingly be computed and
the study of the $\Delta I = 1/2$ rule performed.

The first ingredient that makes the predictions of the $\chi$QM 
substantially different from the
VSA results 
is the inclusion of  corrections of order  $O (\alpha_s N_c)$ coming 
from the
non-factorizable gluon condensates,
first considered by
Pich--de Rafael~\cite{PdeR} (see ref.~\cite{PJ} for an updating
 of this analysis). These are truly non-perturbative effects
which represent a crucial step toward the understanding of the selection rule.
Their most relevant effect is to give the required
 suppression of the $A_2$ amplitude.

To the amplitudes thus obtained, we have consistently
added the renormalization induced
by meson loops, that we have computed in I. 
This gives the needed long-distance
enhancement of $A_0$ while little affecting $A_2$. 
The meson-loop renormalization also provides
the 
scale dependence of the hadronic matrix elements which is necessary in order to
make the matching with
 the Wilson coefficients consistent.

The idea of matching the scale dependence of 
the meson renormalization to the short-distance 
$\mu$-dependence of the Wilson coefficients goes back to the work of
Bardeen--Buras--Gerard~\cite{BBG} (and subsequently~\cite{Paschos}), 
who, however, applied it in the
framework of a cut-off dependent regularization and therefore with results
that are different  from ours.

All the relevant matrix elements are given in I, to which we refer. 
We report in section 3 those that are most important for the
present analysis. In section 4
we discuss our results at the varying of $\Lambda_{\rm QCD}$, the matching
scale $\mu$ and of the input parameters: gluon and quark
condensates. The $\chi$QM leaves 
us with a free parameter: the constituent quark
mass $M$. In order to restrict the possible values it can take, we require, as 
an additional constraint, that the $\gamma_5$-scheme dependence
of the hadronic matrix elements
compensates maximally that of the Wilson coefficients.

As depicted in fig. 9 of section 5,
the combined effects of all contributions thus included
add up to provide, to a very good 
approximation, the experimental values for the $I=0$ and 2 amplitudes. 
We think that our
analysis shows that the $\Delta I = 1/2$ selection rule is
fully accounted for within the standard model and by our present
understanding of its non-perturbative aspects.

This said, a word of caution  is perhaps advisable. Various
uncertainties related to the input parameters
are necessarily present in our computation and we will
discuss them as they occur. In addition, we have to consider the
 approximations inherent to our approach. In particular, 
 higher-order 
terms $O(p^4)$ in the chiral expansion may
cause a 20-30\% correction~\cite{BEF}. 
This is the systematic uncertainty we 
ascribe to our results.

\section{Effective Quark Lagrangian and NLO Wilson Coefficients}

 The quark effective lagrangian at a scale $\mu < m_c$ can be written
as~\cite{GW}
 \beq
{\cal L}_{\Delta S = 1} = -
\frac{G_F}{\sqrt{2}} V_{ud}\,V^*_{us} \sum_i \Bigl[
z_i(\mu) + \tau y_i(\mu) \Bigr] Q_i (\mu) 
 \, . \label{ham}
\eeq

The $Q_i$ are four-quark operators obtained by integrating out in the standard
model the vector bosons and the heavy quarks $t,\,b$ and $c$. A convenient
and by now standard
basis includes the following ten quark operators:
 \beq
\begin{array}{rcl}
Q_{1} & = & \left( \overline{s}_{\alpha} u_{\beta}  \right)_{\rm V-A}
            \left( \overline{u}_{\beta}  d_{\alpha} \right)_{\rm V-A}
\, , \\[1ex]
Q_{2} & = & \left( \overline{s} u \right)_{\rm V-A}
            \left( \overline{u} d \right)_{\rm V-A}
\, , \\[1ex]
Q_{3,5} & = & \left( \overline{s} d \right)_{\rm V-A}
   \sum_{q} \left( \overline{q} q \right)_{\rm V\mp A}
\, , \\[1ex]
Q_{4,6} & = & \left( \overline{s}_{\alpha} d_{\beta}  \right)_{\rm V-A}
   \sum_{q} ( \overline{q}_{\beta}  q_{\alpha} )_{\rm V\mp A}
\, , \\[1ex]
Q_{7,9} & = & \frac{3}{2} \left( \overline{s} d \right)_{\rm V-A}
         \sum_{q} \hat{e}_q \left( \overline{q} q \right)_{\rm V\pm A}
\, , \\[1ex]
Q_{8,10} & = & \frac{3}{2} \left( \overline{s}_{\alpha} 
                                                 d_{\beta} \right)_{\rm V-A}
     \sum_{q} \hat{e}_q ( \overline{q}_{\beta}  q_{\alpha})_{\rm V\pm A}
\, , 
\end{array}  
\label{Q1-10} 
\eeq
where $\alpha$, $\beta$ denote color indices ($\alpha,\beta
=1,\ldots,N_c$) and $\hat{e}_q$  are quark charges. Color
indices for the color singlet operators are omitted. 
The subscripts $(V\pm A)$ refer to
$\gamma_{\mu} (1 \pm \gamma_5)$.
We recall that
$Q_{1,2}$ stand for the $W$-induced current--current
operators, $Q_{3-6}$ for the
QCD penguin operators and $Q_{7-10}$ for the electroweak penguin (and box)
ones. 

The functions $z_i(\mu)$ and $y_i(\mu)$ are the
 Wilson coefficients and $V_{ij}$ the
Koba\-ya\-shi-Mas\-kawa (KM) matrix elements; $\tau = - V_{td}
V_{ts}^{*}/V_{ud} 
V_{us}^{*}$. 
The numerical values of the Wilson coefficients depend on $\alpha_s$. A  
recent determination~\cite{alfa} at LEP and SLC gives
\beq
\alpha_s (m_Z) = 0.119 \pm 0.006 \, ,
\eeq
which corresponds to
\beq
\Lambda^{(4)}_{QCD} = 350 \pm 100 \: \mbox{MeV} \, .
\label{lambdone}
\eeq
The range in \eq{lambdone} will be used for our numerical estimate of the
amplitudes $A_0$ and $A_2$.

Even though not all the operators in \eq{Q1-10} are independent, this basis
is of particular interest for 
the present numerical analysis because it is that employed
for the calculation of the Wilson coefficients 
to the NLO order in $\alpha_s$~\cite{Monaco,Roma}.

In tables 1 and 2 we give explicitly the Wilson coefficients of the ten
operators at the scale $\mu = 1$ GeV and $\mu = 0.8$ GeV, respectively,
in the naive dimensional regularization (NDR) and  't Hooft-Veltman (HV) 
$\gamma_5$-schemes.
Since $\Re \tau = O(10^{-3})$, the $CP$-conserving component of
$A_0$ and $A_2$ 
are  controlled by the  coefficients $z_i(\mu)$, which do not depend
on $m_t$. 
\begin{table}
\begin{center}
\begin{tabular}{|c|r r||r r||r r|}
\hline
$\Lambda_{QCD}^{(4)}$ & \multicolumn{2}{c||}{ 250 MeV }
                      & \multicolumn{2}{c||}{ 350 MeV } 
                      & \multicolumn{2}{c| }{ 450 MeV } \\
\hline
$\alpha_s(m_Z)_{\overline{MS}}$ 
                      & \multicolumn{2}{c||}{ 0.113 }
                      & \multicolumn{2}{c||}{ 0.119 } 
                      & \multicolumn{2}{c| }{ 0.125 } \\
\hline
\multicolumn{7}{c}{\mbox{HV}}\\
\hline
$z_1$&$(0.0320)$&$-0.539$&$(0.0339)$&$-0.683$&$(0.0355)$&$-0.884$ \\
\hline
$z_2$&$(0.988)$&$1.30$&$(0.987)$&$1.40$&$(0.987)$&$1.56$ \\
\hline
$z_3$&$$&$0.0060$&$  $&$0.0108$&$ $&$0.0210$ \\
\hline
$z_4$&$$&$-0.0142$&$  $&$-0.0230$&$ $&$-0.0389$ \\
\hline
$z_5$&$$&$0.0037$&$  $&$0.0052$&$ $&$0.0068$ \\
\hline
$z_6$&$$&$-0.0128$&$  $&$-0.0204$&$ $&$-0.0341$ \\
\hline
$z_7/\alpha$&$$&$-0.0041$&$  $&$-0.0024$&$ $&$-0.0012$ \\
\hline
$z_8/\alpha$&$$&$0.0087$&$  $&$0.0140$&$ $&$0.0234$ \\
\hline
$z_9/\alpha$&$$&$0.0016$&$  $&$0.0068$&$ $&$0.0140$ \\
\hline
$z_{10}/\alpha$&$$&$-0.0084$&$  $&$-0.0135$&$ $&$-0.0223$ \\
\hline 
\multicolumn{7}{c}{\mbox{NDR}}\\
\hline
$z_1$&$(0.0503)$&$-0.440$&$(0.0533)$&$-0.535$&$(0.0557)$&$-0.644$ \\
\hline
$z_2$&$(0.982)$&$1.23$&$(0.981)$&$1.30$&$(0.980)$&$1.37$ \\
\hline
$z_3$&$$&$0.0100$&$  $&$0.0161$&$ $&$0.0283$ \\
\hline
$z_4$&$$&$-0.0281$&$  $&$-0.0432$&$ $&$-0.0703$ \\
\hline
$z_5$&$$&$0.0067$&$  $&$0.0083$&$ $&$0.0088$ \\
\hline
$z_6$&$$&$-0.0281$&$  $&$-0.0438$&$ $&$-0.0730$ \\
\hline
$z_7/\alpha$&$$&$0.0039$&$  $&$0.0122$&$ $&$0.0218$ \\
\hline
$z_8/\alpha$&$$&$0.0120$&$  $&$0.0208$&$ $&$0.0378$ \\
\hline
$z_9/\alpha$&$$&$0.0102$&$  $&$0.0223$&$ $&$0.0382$ \\
\hline
$z_{10}/\alpha$&$$&$-0.0076$&$  $&$-0.0116$&$ $&$-0.0180$ \\
\hline
\end{tabular}
\end{center}
\caption{NLO Wilson coefficients at $\mu=1$ GeV in the HV and NDR schemes.
The corresponding values at $\mu=m_W$ are given in parenthesis 
($\alpha=1/128$).
In the HV scheme one has $z_{3-10}(m_c)=0$. The coefficients
$z_i(\mu)$, relevant for the study of $CP$ conserving
amplitudes, do not depend on $m_t$.}
\end{table}
\begin{table}
\begin{center}
\begin{tabular}{|c|r r||r r||r r|}
\hline
$\Lambda_{QCD}^{(4)}$ & \multicolumn{2}{c||}{ 250 MeV }
                      & \multicolumn{2}{c||}{ 350 MeV } 
                      & \multicolumn{2}{c| }{ 450 MeV } \\
\hline
$\alpha_s(m_Z)_{\overline{MS}}$ 
                      & \multicolumn{2}{c||}{ 0.113 }
                      & \multicolumn{2}{c||}{ 0.119 } 
                      & \multicolumn{2}{c| }{ 0.125 } \\
\hline
\multicolumn{7}{c}{\mbox{HV}}\\
\hline
$z_1$&$(0.0320)$&$-0.657$&$(0.0339)$&$-0.910$&$(0.0355)$&$-1.36$ \\
\hline
$z_2$&$(0.988)$&$1.38$&$(0.987)$&$1.58$&$(0.987)$&$1.96$ \\
\hline
$z_3$&$$&$0.0137$&$  $&$0.0301$&$ $&$0.0798$ \\
\hline
$z_4$&$$&$-0.0292$&$  $&$-0.0540$&$ $&$-0.115$ \\
\hline
$z_5$&$$&$0.0070$&$  $&$0.0100$&$ $&$0.0123$ \\
\hline
$z_6$&$$&$-0.0275$&$  $&$-0.0515$&$ $&$-0.112$ \\
\hline
$z_7/\alpha$&$$&$-0.0055$&$  $&$-0.0030$&$ $&$-0.0065$ \\
\hline
$z_8/\alpha$&$$&$0.0198$&$  $&$0.0379$&$ $&$0.0827$ \\
\hline
$z_9/\alpha$&$$&$0.0070$&$  $&$0.0203$&$ $&$0.0415$ \\
\hline
$z_{10}/\alpha$&$$&$-0.0181$&$  $&$-0.0330$&$ $&$-0.0644$ \\
\hline 
\multicolumn{7}{c}{\mbox{NDR}}\\
\hline
$z_1$&$(0.0503)$&$-0.524$&$(0.0533)$&$-0.663$&$(0.0557)$&$-0.781$ \\
\hline
$z_2$&$(0.982)$&$1.29$&$(0.981)$&$1.39$&$(0.980)$&$1.48$ \\
\hline
$z_3$&$$&$0.0180$&$  $&$0.0360$&$ $&$0.0870$ \\
\hline
$z_4$&$$&$-0.0471$&$  $&$-0.0852$&$ $&$-0.182$ \\
\hline
$z_5$&$$&$0.0085$&$  $&$0.0077$&$ $&$-0.0129$ \\
\hline
$z_6$&$$&$-0.0495$&$  $&$-0.0947$&$ $&$-0.226$ \\
\hline
$z_7/\alpha$&$$&$0.0073$&$  $&$0.0204$&$ $&$0.0366$ \\
\hline
$z_8/\alpha$&$$&$0.0280$&$  $&$0.0589$&$ $&$0.143$ \\
\hline
$z_9/\alpha$&$$&$0.0206$&$  $&$0.0441$&$ $&$0.0779$ \\
\hline
$z_{10}/\alpha$&$$&$-0.0159$&$  $&$-0.0267$&$ $&$-0.0438$ \\
\hline
\end{tabular}
\end{center}
\caption{Same as in table 1 at $\mu=0.8$ GeV.}
\end{table}

\section{The Hadronic Matrix Elements}

 The $\chi$QM allows us to compute hadronic matrix elements by
coupling quarks to the chiral Goldstone bosons. The model 
and the details of the derivation of the matrix elements are 
discussed in I, to which we refer the interested reader.

The results for the  matrix elements 
\beq
\langle Q_i \rangle _{0,2} \equiv
\langle 2 \, \pi \, , I=0,2 \, | Q_i | K^0 \rangle
\eeq
are given by
the equations (8.13)-(8.48) of I, including all contributions of
order $O (N^2)$, $O (N)$ and $O (\alpha_s N)$.
We have used dimensional regularization and, in dealing with
the $\gamma_5$ matrix, we have worked out the analysis
 in both the HV and the NDR schemes. 

For reference, we report here those elements that are
most relevant to the present discussion. The contributions to $A_0$ and $A_2$
of the electroweak penguin operators $Q_{7-10}$ are suppressed
by the smallness of their Wilson coefficients.

In the HV scheme we find:
\bea
\langle Q_1 \rangle _0 & = & \frac{1}{3} X \left[ -1 + \frac{2}{N_c} \left(
1 - \delta_{\vev{GG}} \right)
\right] + a_0 (Q_1)\\
\langle Q_1 \rangle _2 & = & \frac{\sqrt{2}}{3} X \left[ 1 + \frac{1}{N_c} 
\left(
1 - \delta_{\vev{GG}} \right) 
\right] + a_2 (Q_1)\\
\langle Q_2 \rangle _0 & = & \frac{1}{3} X \left[ 2  - \frac{1}{N_c} \left(
1 - \delta_{\vev{GG}} \right) 
\right] + a_0 (Q_2)\\
\langle Q_2 \rangle _2 & = &  \frac{\sqrt{2}}{3} X \left[ 1 + \frac{1}{N_c} 
\left( 1 - \delta_{\vev{GG}} \right) \right] + a_2 (Q_2)\\
\langle Q_3 \rangle _0 & = & \frac{1}{N_c} X  \left(
1 - \delta_{\vev{GG}} \right) + a_0 (Q_3)\\
\langle Q_4 \rangle _0 & = & X + a_0 (Q_4) \\
\langle Q_5 \rangle _0 & = &  \frac{2}{N_c}  \, 
\frac{\langle \bar{q}q \rangle}{M f_\pi^2} \, X'  + a_0 (Q_5)\\
\langle Q_6 \rangle _0 & = & 2  \, \frac{\langle 
\bar{q}q \rangle}{f_\pi^2 M} \, X'  + a_0 (Q_6) 
\eea
where
\beq
X \equiv \sqrt{3} f_\pi \left( m_K^2 - m_\pi^2 \right) \quad
\mbox{and} \quad X' \equiv X \left( 1 - 6\ \frac{M^2}{\Lambda_\chi^2} \right) 
\, . \label{x'}
\eeq

In the NDR:
 \bea
\langle Q_1 \rangle _0 & = & \frac{1}{3} X \left[ -1 + \frac{2}{N_c} \left(
1 - \delta_{\vev{GG}} \right)
\right] + a_0 (Q_1)\\
\langle Q_1 \rangle _2 & = & \frac{\sqrt{2}}{3} X \left[ 1 + \frac{1}{N_c} 
\left(
1 - \delta_{\vev{GG}} \right) 
\right] + a_2 (Q_1)\\
\langle Q_2 \rangle _0 & = & \frac{1}{3} X \left[ 2  - \frac{1}{N_c} \left(
1 - \delta_{\vev{GG}} \right) 
\right] + a_0 (Q_2)\\
\langle Q_2 \rangle _2 & = &  \frac{\sqrt{2}}{3} X \left[ 1 + \frac{1}{N_c} 
\left( 1 - \delta_{\vev{GG}} \right) \right] + a_2 (Q_2)\\
\langle Q_3 \rangle _0 & = & \frac{1}{N_c} \left( X'
-  \delta_{\vev{GG}} X \right) + a_0 (Q_3)
\\
\langle Q_4 \rangle _0 & = & X'  +a_0 (Q_4)\\
\langle Q_5 \rangle _0 & = & \frac{2}{N_c}  \,
\frac{\langle \bar{q}q \rangle}{M f_\pi^2} \, X''  + a_0 (Q_5)  \\
\langle Q_6 \rangle _0 & = & 2   \, \frac{\langle 
\bar{q}q \rangle }{f_\pi^2 M} \, X''  +a_0 (Q_6)  
\eea
where
\beq
X''  =   X \left( 1 - 9\, 
\frac{M^2}{\Lambda_\chi^2} \right) \, . \label{x''}
\eeq

The function $a_{0,2}(Q_i)$ represent the one-loop mesonic renormalization of
the hadronic elements and they are given 
in I.   They are made of polynomial terms, generally
of the order of
 \beq
 \frac{m^2}{(4 \pi f)^2} \, ,
 \eeq
and logarithmic terms of the order of
 \beq
 \frac{m^2}{(4 \pi f)^2} \ln \frac{m^2_a}{m^2_b} \, ,
 \eeq
where the masses can be any among $m_\pi$, $m_K$ and $m_\eta$ and $f$ is
the pion decay constant in the tree-level chiral lagrangian.

  The one-loop 
renormalization of $f$  is taken into account by replacing
$f$ with $f_1$ in
the tree-level amplitudes, which amounts to replacing $1/f^3$ with $1/f^3_\pi$
multiplied by
\beq
1 + 3\ \frac{f_\pi - f_1}{f_\pi} \simeq 1.18\, .
\eeq

The corrections of $O (\alpha_s N)$ are important.
They are parametrized by the value of the gluonic condensate: 
\beq
 \delta_{\langle GG \rangle} = \frac{N_c}{2} \frac{\langle 
 \alpha_s G G/\pi \rangle}{16 \pi^2 f^4} \, . \label{GG}
\eeq
Their most important effect is to reduce the contribution of the 
operator $Q_1$ and $Q_2$ to the amplitude $A_2$; in fact, 
\beq
A_2 \approx ( z_1 + z_2 ) \left[ 1 + \frac{1}{N_c} ( 1 - 
 \delta_{\langle GG \rangle} ) \right]
 \eeq
and by taking $\delta_{\langle GG \rangle} \simeq 3$, which is what we obtain
for the central value
of the gluon condensate (see \eq{GGexp} below) , the correction is
 large enough to
revert the sign of the $1/N_c$ term and thus suppress the amplitude.

A relevant contribution to
the amplitude $A_0$ arises from
the gluonic penguins $Q_{5,6}$, whose matrix elements are directly 
proportional to
the value of $\vev{\bar{q}q}$ and are controlled
by $M$ through the suppression factors 
in (\ref{x'}) and
(\ref{x''}), that
make the matrix elements larger for smaller values of $M$.

The meson-loop renormalization is sizable. 
It enhances $A_0$ while little affecting $A_2$, as it is needed in order to
reproduce the experimental values.
An important feature of this correction is
 the additional scale dependence that is introduced in 
the hadronic matrix elements and that  matches to
a good approximation that of the Wilson 
coefficients.

For the purpose of comparison with the existing literature, it is useful
to introduce the effective factors
\beq
B_i^{(0,2)} \equiv \frac{\langle Q_i \rangle _{0,2}^{\chi {\rm QM}}}
{\langle Q_i \rangle _{0,2}^{\rm VSA}} \, ,
\eeq
that give the ratio between our hadronic matrix elements and those of the
VSA.
We shall discuss their numerical values in section 4.3. 

\subsection{Input Parameters}

 The quark and the gluon condensates are two  input parameters of
 our computation. As discussed in I,
 their phenomenological determination  
is a complicated 
question (they parametrize the genuine non-perturbative part of the 
computation) and the literature offers different estimates. 

We  identify the condensates entering our computation with those obtained by
fitting the experimental data  by means of the QCD sum 
rules (QCD--SR) or
lattice computations.
In our discussion we will vary
these input parameters within the given bounds and
obtain a range of values for the amplitudes we are interested in.

A review of recent determinations of these parameters, together
 with a justification of the estimated errors, is given in I. 
Here we only report the ranges that we will explore in our
numerical analysis.

For the gluon condensate, we take 
the scale independent range
\beq
\langle \frac{\alpha_s}{\pi} G G \rangle = (376 \pm  47 \:
\mbox{MeV} )^4 \, , \label{GGexp}
\eeq
which encompasses the results of recent QCD-SR
analysis~\cite{Narison}. 

For the quark condensate, we consider the range
\beq
 - ( 200 \: \mbox{MeV} )^3 \leq 
 \vev{\bar{q}q} \leq  - ( 280 \: \mbox{MeV} )^3 \label{range}
\label{qqexp}
\eeq 
in order to include the central values and the errors
of the  QCD-SR~\cite{DdeR} and lattice estimates~\cite{qqlattice1}.

In discussing the scale dependence of our results,
it is necessary to include the perturbative running of the quark
condensate. This can be done in the QCD-SR approach by
 using the
renormalization-group  running masses $\overline{m}_u + \overline{m}_d$, the
value of
which is estimated at $\mu = 1$ GeV to be~\cite{BPdeR}
\beq
\overline{m}_u + \overline{m}_d = 12 \pm 2.5 \: \mbox{MeV} \label{mumd}
\eeq
for $\Lambda^{(3)}_{\rm QCD} = 300 \pm 150$ MeV. The error
in
(\ref{mumd}) reflects changes in the spectral functions. 
In our numerical estimates,  we will take as input values the running masses at
1 GeV given by (\ref{mumd}). Even though our preferred range of
$\Lambda^{(4)}_{\rm QCD}$ in eq. (\ref{lambdone}) corresponds to
$\Lambda^{(3)}_{\rm QCD} = 400 \pm 100 \: \mbox{MeV}$,
 and therefore is not that employed in ref.~\cite{BPdeR}, we feel that we are
not making too large an error since the determination in \eq{mumd} is not very
sensitive to the choice of $\Lambda_{\rm QCD}$.

By taking the value (\ref{mumd}),
we find  for the scale-dependent (and normal-ordered)
condensate
\beq
\vev{\bar{q}q} (\mu) = - \frac{f_\pi^2 m_\pi^2 ( 1 - 
\delta_\pi)}{\overline{m}_u(\mu) + \overline{m}_d (\mu)} \, , \label{qqQCD}
\eeq\
 corresponding ($\delta_\pi$ is  a few percent correction) 
to the numerical values of
\beq
\vev{\bar{q}q} = - (238 \pm 19 \: \mbox{MeV} )^3 
\label{qq1}
\eeq
at 1 GeV and
\beq
\vev{\bar{q}q} = - (222 \pm 19 \: \mbox{MeV} )^3
\label{qq2}
\eeq
at 0.8 GeV.
The error in \eqs{qq1}{qq2} corresponds to that in (\ref{mumd}). 
The central values
in \eqs{qq1}{qq2} are in the lower half of the range (\ref{range}).

\section{Computing $A_0$ and $A_2$}

The physics 
entering the $\Delta I = 1/2$ rule involves very different scales and thus 
different effective theories. As a consequence, any attempt to an
explanation must
necessarily
depends on several parameters.

Our solution has two input parameters, 
the quark and gluon condensates, as well 
as a free,
model-dependent parameter: the constituent quark mass $M$. 
In addition, the short-distance part depends on 
several high-energy parameters like the KM matrix elements and 
the masses of the heavy quarks. 

We include all ten operators (\ref{Q1-10}) even though the effect of
the electroweak operators is only of a few percents.
We proceed step-by-step summarizing our result 
by means of tables and figures. 

\subsection{The VSA Approach}

 First of all, in order to make possible  gauging
  the progress of our computation of the hadronic matrix 
elements, 
 we report in table 3 the
results one would obtain by using
the VSA  for the evaluation of the hadronic matrix elements.

\begin{table}
\begin{small}
\begin{center}
\begin{tabular}{|c||c|c||c|c||c|c|}
\hline
 & \multicolumn{2}{c||}{$\mu = 0.8$ GeV} & \multicolumn{2}{c||}{$\mu = 0.9$ GeV}
 & \multicolumn{2}{c|}{$\mu = 1$ GeV} \\
\hline
       & NDR & HV & NDR & HV & NDR & HV \\
\hline
$A_0$  & 1.11 & 0.93 & 0.89 & 0.77 & 0.75 & 0.67  \\
\hline
$A_2$  & 3.00 & 2.76 & 3.07 & 2.87 & 3.13 & 2.95  \\
\hline
$\Delta_{\gamma_5} A_0$ & \multicolumn{2}{c||}{$18 \%$ } & \multicolumn{2}{c||}{$15 \%$ } &
\multicolumn{2}{c|}{$12 \%$ }\\
\hline
$\Delta_{\gamma_5} A_2$ & \multicolumn{2}{c||}{$8 \%$ } & \multicolumn{2}{c||}{$7 \%$ } &
\multicolumn{2}{c|}{$6 \%$ }\\
\hline
$\Delta_\mu A_0$ & \multicolumn{6}{c|}{$39\% - 33 \%$ } \\
\hline
$\Delta_\mu A_2$ & \multicolumn{6}{c|}{$4\% - 7 \%$ } \\
\hline
\end{tabular}
\end{center}
\end{small}
\caption{Matching-scale and $\gamma_5$-scheme dependence 
of $A_0$ (in units of $10^{-7}$ GeV) and $A_2$
(in units of $10^{-8}$ GeV) in the VSA approach with NLO Wilson coefficients. 
The amplitudes are computed for 
$\Lambda^{(4)}_{\rm QCD}$  = 350 MeV and 
$\vev{\bar{q}q}$ given in \eq{qqQCD}. The two values quoted for the
$\mu$-dependence of the amplitudes correspond to the NDR and HV scheme results
 in the range between 0.8 and 1.0 GeV.}
\end{table}

Several comments are in order. At $\mu = 0.8$~GeV $\simeq \Lambda_{\chi}$
(that is our preferred matching
scale) and both in the NDR and the HV
schemes, $A_0$ is too small by a factor of three while 
$A_2$ is two times too large. Therefore, 
the ratio 
\beq
\omega^{-1} \equiv \Re A_0/\Re A_2 \simeq 3.7 \, ,
\eeq
is about a
factor six smaller than the experimental value 
$\omega^{-1} = 22.2$.

This estimate is done by using the hadronic matrix element
for the penguin operators  given by the VSA. We have
used for the quark condensate the PCAC expression given in \eq{qqQCD}. 
As we shall see in the following,
a  larger value for $A_0$ can be obtained by using
larger values for the quark condensate. A popular choice is
\beq
\vev{\bar{q}q} ( 1 \mbox{GeV} ) =
\left( - \frac{m_K^2 f_K^2}{m_s (1 \mbox{GeV})} \right) ^{1/3}  
\simeq - 260 \: \mbox{MeV} \, .
\eeq

The literature also offers determinations of the selection rule 
in which the matching between Wilson coefficients and hadronic matrix 
elements is
done without meson-loop renormalization but at a
 much lower values of matching 
(at about 300 MeV, for instance \cite{PdeR}). Such a procedure 
is no longer justifiable in view of the NLO determination of the Wilson
coefficients that shows, for such a low-energy matching, the breaking down
of the perturbative expansion.

 In table 3 we have defined
\beq
\Delta_\mu A_i \equiv 2 \:\left| \frac{A_i (0.8 \: \mbox{GeV}) -
 A_i (1.0 \: \mbox{GeV})}
 {A_i ( 0.8 \: \mbox{GeV}) + A_i(1.0 \: \mbox{GeV}) }\right| \, ,
 \eeq
 as a direct measure of the scale dependence.
 The scale 
dependence of $A_0$  is rather large, with
 $\Delta_\mu A_0$  almost 40\% in the NDR scheme. 
 On the other hand, the $A_2$'s scale dependence
 remains within 10\%.

The difference between the HV and NDR results is quantified by
\beq
\Delta_{\gamma_5} A_i
 \equiv 2 \:\left| \frac{A_i^{\rm NDR} - A_i^{\rm HV}}{A_i^{\rm NDR} + 
 A_i^{\rm HV}}\right| \, .
 \eeq
Contrary to the scale dependence, 
the
$\gamma_5$-scheme
dependence  is not large for the amplitude $A_0$ and $A_2$,
 remaining below 10\% and it is not an issue in this paper. 
 It is however very large in other observable
 quantities like, for instance, $\varepsilon '/\varepsilon$ where it 
 reaches in the $1/N_c$ approach the 80\% level. 
 
 Because the VSA hadronic matrix elements  depend neither on the 
 scale\fnote{\dag}{Except for the perturbative running of the quark condensate.}
 nor  the $\gamma_5$-scheme, the
  dependences in table 3 are a direct measure of those of
the Wilson coefficients.

We now turn to the $\chi$QM model determination of the selection rule.

\subsection{Matching and scale dependence}

In matching the short-distance Wilson coefficients to the hadronic matrix 
elements, we combine terms that are scale dependent. Ideally, the scale 
dependence of the Wilson coefficients should compensate against that of the 
hadronic matrix elements to provide a scale 
independent result. 
In practice, within our approximations and for the central value of
$\Lambda^{(4)}_{\rm QCD}$, we find that the scale dependence in 
our matching remains below 20\%. In particular, as 
shown in table 4, the difference in the value of $A_0$ between the  matching  at 1 GeV
and that at $0.8$ is about 20\% in both the HV and NDR, to be compared
with the values given in table 3, where it is about 30\%.
 For $A_2$ it is respectively 12 and 1\%.
\begin{table}
\begin{small}
\begin{center}
\begin{tabular}{|c||c|c||c|c||c|c|}
\hline
\multicolumn{7}{|c|}{$\Lambda^{(4)}_{\rm QCD}$ = 250 MeV}\\
\hline 
 & \multicolumn{2}{c||}{$\mu = 0.8$ GeV} & \multicolumn{2}{c||}{$\mu = 0.9$ GeV}
 & \multicolumn{2}{c|}{$\mu = 1$ GeV} \\
\hline
       & NDR & HV & NDR & HV & NDR & HV \\
\hline
$A_0$  & 2.31 & 2.29 & 2.17 & 2.15 & 2.08 & 2.04  \\
\hline
$A_2$  & 1.66 & 1.56 & 1.73 & 1.65 & 1.80 & 1.72  \\
\hline
$\Delta_{\gamma_5} A_0$ & \multicolumn{2}{c||}{$ < 1 \%$ } & 
\multicolumn{2}{c||}{$ 1 \%$ } &
\multicolumn{2}{c|}{$2 \%$ }\\
\hline
$\Delta_{\gamma_5} A_2$ & \multicolumn{2}{c||}{$6 \%$ } & 
\multicolumn{2}{c||}{$5 \%$ } &
\multicolumn{2}{c|}{$5 \%$ }\\
\hline
$\Delta_\mu A_0$ & \multicolumn{6}{c|}{$10\% - 12\%$ } \\
\hline
$\Delta_\mu A_2$ & \multicolumn{6}{c|}{$8\% -10\%$ } \\
\hline
\hline
\multicolumn{7}{|c|}{$\Lambda^{(4)}_{\rm QCD}$ = 350 MeV}\\
\hline 
 & \multicolumn{2}{c||}{$\mu = 0.8$ GeV} & \multicolumn{2}{c||}{$\mu = 0.9$ GeV}
 & \multicolumn{2}{c|}{$\mu = 1$ GeV} \\
\hline
       & NDR & HV & NDR & HV & NDR & HV \\
\hline
$A_0$  & 2.97 & 2.94 & 2.66 & 2.61 & 2.45 & 2.39  \\
\hline
$A_2$  & 1.60 & 1.46 & 1.68 & 1.56 & 1.75 & 1.64  \\
\hline
$\Delta_{\gamma_5} A_0$ & \multicolumn{2}{c||}{$ 1 \%$ } & \multicolumn{2}{c||}{$ 2 \%$ } & \multicolumn{2}{c|}{$2 \%$ }\\
\hline
$\Delta_{\gamma_5} A_2$ & \multicolumn{2}{c||}{$9 \%$ } & \multicolumn{2}{c||}{$8 \%$ } &
\multicolumn{2}{c|}{$6 \%$ }\\
\hline
$\Delta_\mu A_0$ & \multicolumn{6}{c|}{$19\% -20\%$ } \\
\hline
$\Delta_\mu A_2$ & \multicolumn{6}{c|}{$9\% - 12\%$ } \\
\hline
\hline
\multicolumn{7}{|c|}{$\Lambda^{(4)}_{\rm QCD}$ = 450 MeV}\\
\hline 
 & \multicolumn{2}{c||}{$\mu = 0.8$ GeV} & \multicolumn{2}{c||}{$\mu = 0.9$ GeV}
 & \multicolumn{2}{c|}{$\mu = 1$ GeV} \\
\hline
       & NDR & HV & NDR & HV & NDR & HV \\
\hline
$A_0$  & 4.26 & 4.17 & 3.52 & 3.41 & 3.06 & 2.95  \\
\hline
$A_2$  & 1.58 & 1.34 & 1.64 & 1.46 & 1.71 & 1.55  \\
\hline
$\Delta_{\gamma_5} A_0$ & \multicolumn{2}{c||}{$ 2 \%$ } & 
\multicolumn{2}{c||}{$ 3 \%$ } &
\multicolumn{2}{c|}{$4 \%$ }\\
\hline
$\Delta_{\gamma_5} A_2$ & \multicolumn{2}{c||}{$17 \%$ } & 
\multicolumn{2}{c||}{$12 \%$ } &
\multicolumn{2}{c|}{$9 \%$ }\\
\hline
$\Delta_\mu A_0$ & \multicolumn{6}{c|}{$33\% - 34 \%$ } \\
\hline
$\Delta_\mu A_2$ & \multicolumn{6}{c|}{$8\% - 15\%$ } \\
\hline
\end{tabular}
\end{center}
\end{small}
\caption{Same as in table 3 in the $\chi$QM approach, for different
values of $\Lambda^{(4)}_{\rm QCD}$. We take for the gluon condensate the
central value $\vev{\alpha_s GG / \pi}$ = (376 MeV)$^4$ and for
the quark
condensate $\vev{\bar{q}q}$  \eq{qqQCD}. The
$\gamma_5$-scheme stability is optimized at $M$ = 160 MeV. 
}
\end{table}

As expected,
the matching dependence is stronger for larger values of $\Lambda_{\rm QCD}$. 
We
take the central value at 350 MeV as our reference value for the
discussion. 

Thanks to the satisfactory scale independence, we can take the 
matching at any values between 0.8 and 1 GeV. Yet it is useful
to bear in mind that
any number quoted in what follows
suffers of an intrinsic uncertainty of at least 20\% because
of the residual scale dependence in the matching itself. 
We take the matching 
at 0.8 GeV as the best compromise between 
the range of validity of the perturbative regime and that of
chiral perturbation theory.

We have also verified the consistency of the perturbative expansion 
in the short-distance regime by comparing
leading order and the NLO results for $A_0$ and $A_2$ at $\mu = 0.8$ GeV. 
The change in $A_0$ turns out to be of
about 10\% and similarly for $A_2$. These  
values confirm that the choice of the matching scale $\mu = 0.8$ GeV is well 
within the
perturbative regime of the QCD short-distance analysis.  

The values for $A_0$ and $A_2$ in table 4 should not be compared directly
to the experimental values because they correspond to a specific
choice for the quark and gluon condensates, as well as of $M$. A
 more complete discussion of the dependence of the
results on the input parameters is presented in 
 section 4.4.
 
\subsection{The $B_i$ Factors}

For the purpose of comparison with the existing literature,
we collect in table 5 the $B_i$ factors for the hadronic matrix elements
of the operators in \eq{Q1-10}. The values of the $B_i$ depend on the
scale at which the matrix elements
are evaluated, on the input parameters,
on $M$, and on the $\gamma_5$-scheme; 
we have given in table 5 a representative example of their size.

Only the first six operators are relevant for our present analysis and we 
postpone any comment on the electroweak penguin matrix elements when
studying $\varepsilon'/\varepsilon$.

The values of $B_1^{(0)}$ and $B_2^{(0)}$ show 
that the corresponding
hadronic matrix elements in the $\chi$QM are, once non-factorizable
contributions and meson renormalization have been included, respectively
twelve and three times
larger than their VSA values. 
At the same time, 
$B_1^{(2)}$ and $B_2^{(2)}$ turn out to be about half of what found in the VSA. 
These features make
it possible for the selection rule to be reproduced in the $\chi$QM.

For comparison, in the $1/N_c$ approach of ref. \cite{BBG}, the inclusion of
meson-loop renormalization through a cutoff regularization leads, at the
scale of 1 GeV, to $B_1^{(0)} = 5.2$, $B_2^{(0)} = 2.2$ and 
$B_1^{(2)} = B_2^{(2)} = 0.55$ (see ref.~\cite{BBH}), 
a result that is not sufficient to
reproduce the $\Delta I = 1/2$ rule.
The value $B_1^{(2)} = B_2^{(2)} = 0.55$ in both 
the $\chi$QM (table 5) and the $1/N_c$ approaches is remarkable, and yet
a numerical coincidence, since the suppression orginates from gluon
condensate corrections in the $\chi$QM, whereas it is the effect of
the meson loop renormalization (regularized via explicit cut-off)
in ref. \cite{BBG}. 

The values of the penguin matrix elements $\vev{Q_3}$ 
and $\vev{Q_4}$ in the $\chi$QM lead to rather large $B_i$ factors. 
In the case of $Q_3$, the $\chi$QM result has the opposite sign of the VSA
result and $B_3$ is negative.
This is the effect of the large non-perturbative gluon
correction.

Regarding the gluon penguin operator $Q_6$ (and $Q_5$), we find that 
the $\chi$QM gives
a result consistent with the VSA (and the $1/N_c$ approach), 
$B_6$ ($B_5$) being approximately equal to two for
small values of the quark condensate and one  at larger values. It is the
quadratic dependence (to be contrasted to the linear dependence in the
$\chi$QM) of the VSA matrix element for the penguin operators that
it responsible for the different weight of these operators at different values
of the quark condensate. The lattice estimate for these operators 
gives $B_5 = B_6 = 1.0 \pm 0.2$~\cite{Martinelli}.

\begin{table}
\begin{small}
\begin{center}
\begin{tabular}{|c||c|c||c|c|}
\hline
 & \multicolumn{2}{|c||}{\rm HV} & \multicolumn{2}{c|}{\rm NDR} \\ 
\cline{2-5}
\cline{2-5}
 & $\mu = 0.8$ GeV & $\mu = 1.0$ GeV & $\mu = 0.8$ GeV & $\mu = 1.0$ GeV\\
\hline
$B^{(0)}_1$  & 10.6 & 11.1 & 10.6 & 11.1\\
\hline
$B^{(0)}_2$  & 2.8 & 3.0  & 2.8 & 3.0 \\
\hline
$B^{(2)}_1$  & 0.52 & 0.55 & 0.52 & 0.55\\
\hline
$B^{(2)}_2$ & 0.52 & 0.55 & 0.52 & 0.55\\
\hline
$B_3$ & $-2.9$ & $-3.0$ & $-3.4$ & $-3.6$\\
\hline
$B_4$ & 1.8 & 1.9 & 1.3 & 1.4\\
\hline
$B_5 = B_6$ & $2.6 \div 0.93$ & $2.7 \div 0.98$ & $2.0 \div 0.74$ & $2.2 \div 0.78$\\
\hline
$B_7^{(0)}$ &$3.2 \div 2.3$ &  $3.4 \div 2.4$ & $3.0 \div 2.2$  & $3.3 \div 2.4$ \\
\hline
$B_8^{(0)}$ &$3.5 \div 2.3$& $3.8 \div 2.5$ &  $3.3 \div 2.2$ & $3.6 \div 2.4$ \\
\hline
$B_9^{(0)}$ &3.9 & 4.0 & 3.6  & 3.8\\
\hline
$B_{10}^{(0)}$ &4.4 & 4.7 & 5.2 & 5.5\\
\hline
$B_7^{(2)}$ &$2.6\div 1.4$ & $2.8 \div 1.5$ & $2.6 \div 1.4$ & $2.8 \div 1.5$\\
\hline
$B_8^{(2)}$ &$2.1\div 1.4$& $2.2 \div 1.5$ & $2.0 \div 1.4$ & $2.2 \div 1.4$\\
\hline
$B_9^{(2)}$ &0.52 & 0.55 & 0.52 & 0.55\\
\hline
$B_{10}^{(2)}$ & 0.52 & 0.55 & 0.52 & 0.55\\
\hline
\end{tabular}
\end{center}
\end{small}
\caption{The $B_i$ factors in the $\chi$QM (including meson-loop
renormalizations) 
at two different scales: $\mu$ = 0.8 and 1.0 GeV. We have taken
the gluon
condensate at the central value of \eq{GGexp}, while the ranges given
for $B_{5-8}$
correspond to varying the quark condensate according to \eq{qqexp}. The 
results shown are given in the HV and NDR schemes  for 
$M = 180$ MeV.}
\end{table}

\clearpage
\subsection{Dependence on the Input Parameters }

In this section we 
study the dependence of $A_0$ and $A_2$ on  the quark
 and gluon condensates.
Figs. 1, 2, 3 and 4 show such a dependence for 
$\Lambda_{\rm QCD}^{(4)} = 350$ MeV 
and four representative values of $M$. 

Each figure contains the HV 
result (black
lines) as well as the NDR one (grey lines). The spread between these two
determinations is mostly due to $A_2$ that contains an irreducible
$\gamma_5$-scheme dependence, as explained in section 4.5.

The gluon condensate 
dependence is represented by the three
horizontal bands, the central one corresponding to the central value
given in \eq{GGexp}, while the other two bound the one standard 
deviation range.
A larger gluon condensate leads to lower values of $A_2$.

The dependence on the quark condensate is represented by the points on a given 
line. The lenght of the lines corresponds
to the range given in \eq{qqexp}. 

Larger values of $A_0$ are obtained for larger values of the
quark condensate and smaller values of $M$. This is  the
effect of the gluon penguin operators $Q_{5,6}$, 
whose matrix elements are proportional to the
factor
\beq
\vev{\bar q q}\left( 1 - 6 \, \frac{M^2}{\Lambda_\chi^2} \right)
\eeq
in the HV scheme, and similarly (with 6 being replaced by 9) 
in the NDR scheme. 
\begin{figure}[ht]
\epsfxsize=12cm
\centerline{\epsfbox{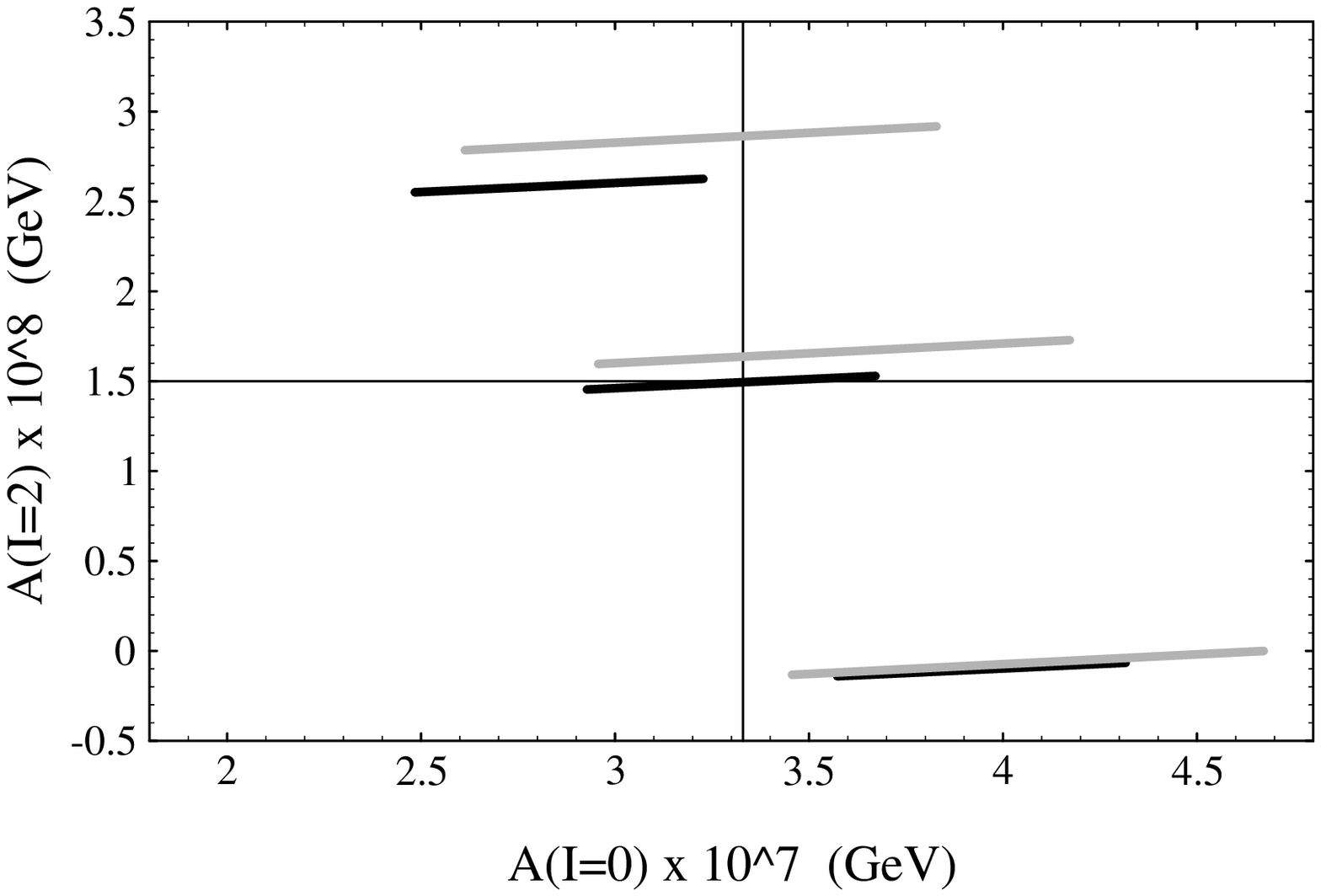}}
\caption{Dependence of $A_0$ and $A_2$ on
$\vev{\bar{q}q}$ and $\vev{GG}$ for $\Lambda^{(4)}_{\rm QCD} = 0.350$ GeV,
$\mu = 0.8$ GeV,
and $M = 160$ MeV. 
The black (grey) lines represent the HV (NDR) results
by varying $\vev{\bar{q}q}$ in the range of \eq{qqexp} for fixed
$\vev{\alpha_s GG/\pi}$. The three bands
correspond to varying $\vev{\alpha_s GG/\pi}$ in the range 
of \eq{GGexp}, with central
lines correponding to the central value of $\vev{GG}$. 
The experimental values of $A_0$ and $A_2$ are given by the cross hairs. 
The small dependence of $A_2$ on the quark condensate is due to the 
 the contribution of the electroweak penguins $Q_{7,8}$ .}
\end{figure}
\begin{figure}
\epsfxsize=12cm
\centerline{\epsfbox{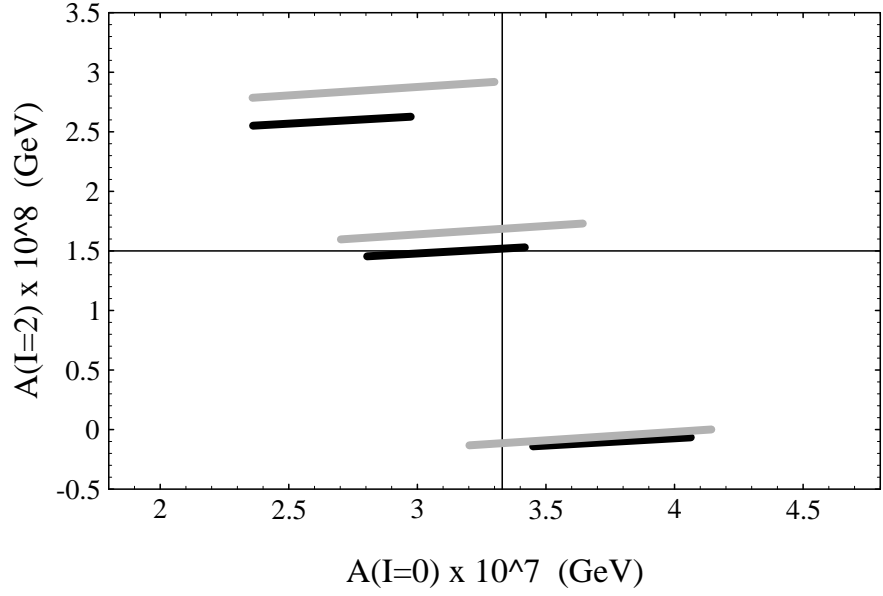}}
\caption{Same as in fig. 1 for $M = 180$ MeV.}
\end{figure}
\begin{figure}
\epsfxsize=12cm
\centerline{\epsfbox{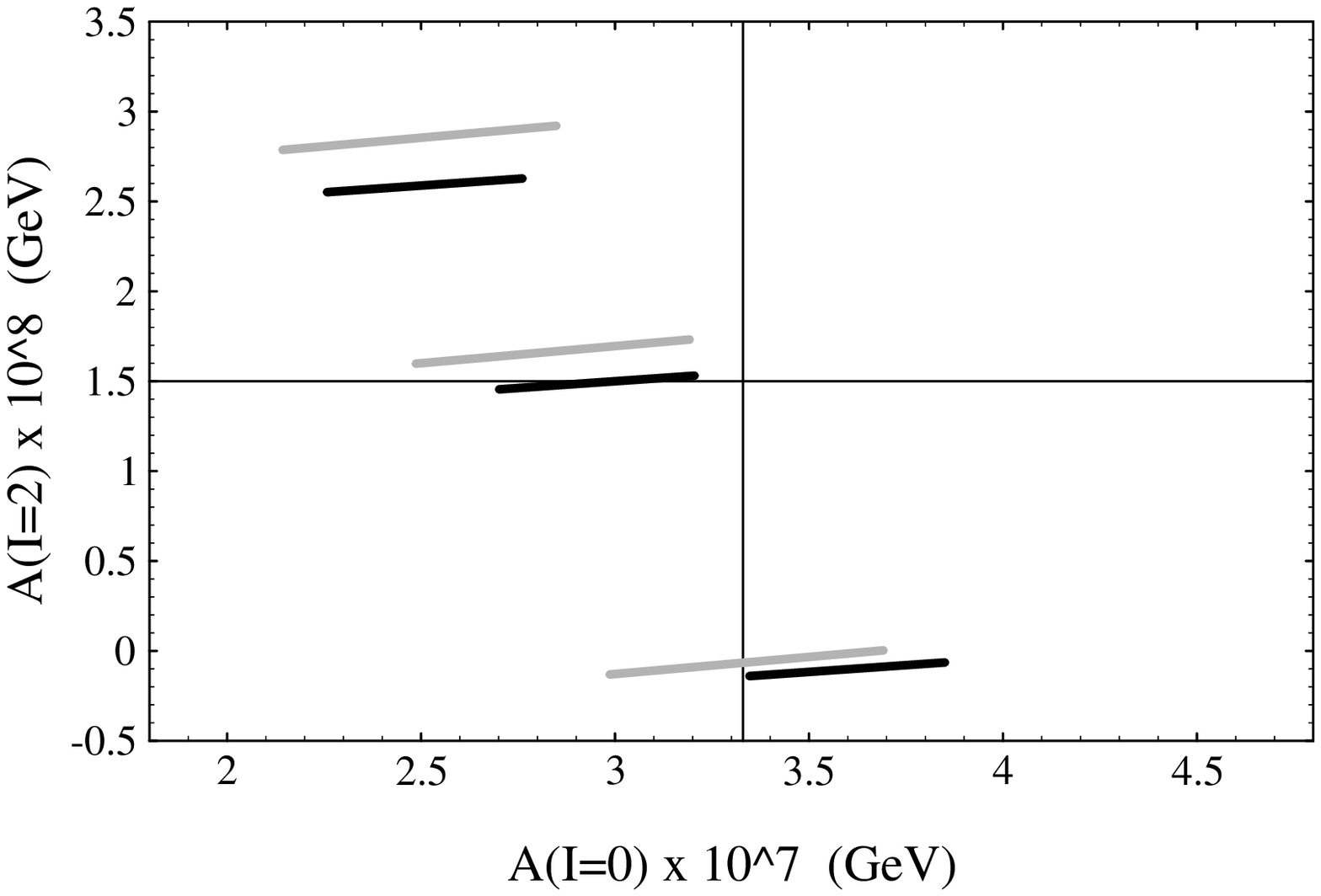}}
\caption{Same as in fig. 1 for $M = 200$ MeV.}
\end{figure}
\begin{figure}
\epsfxsize=12cm
\centerline{\epsfbox{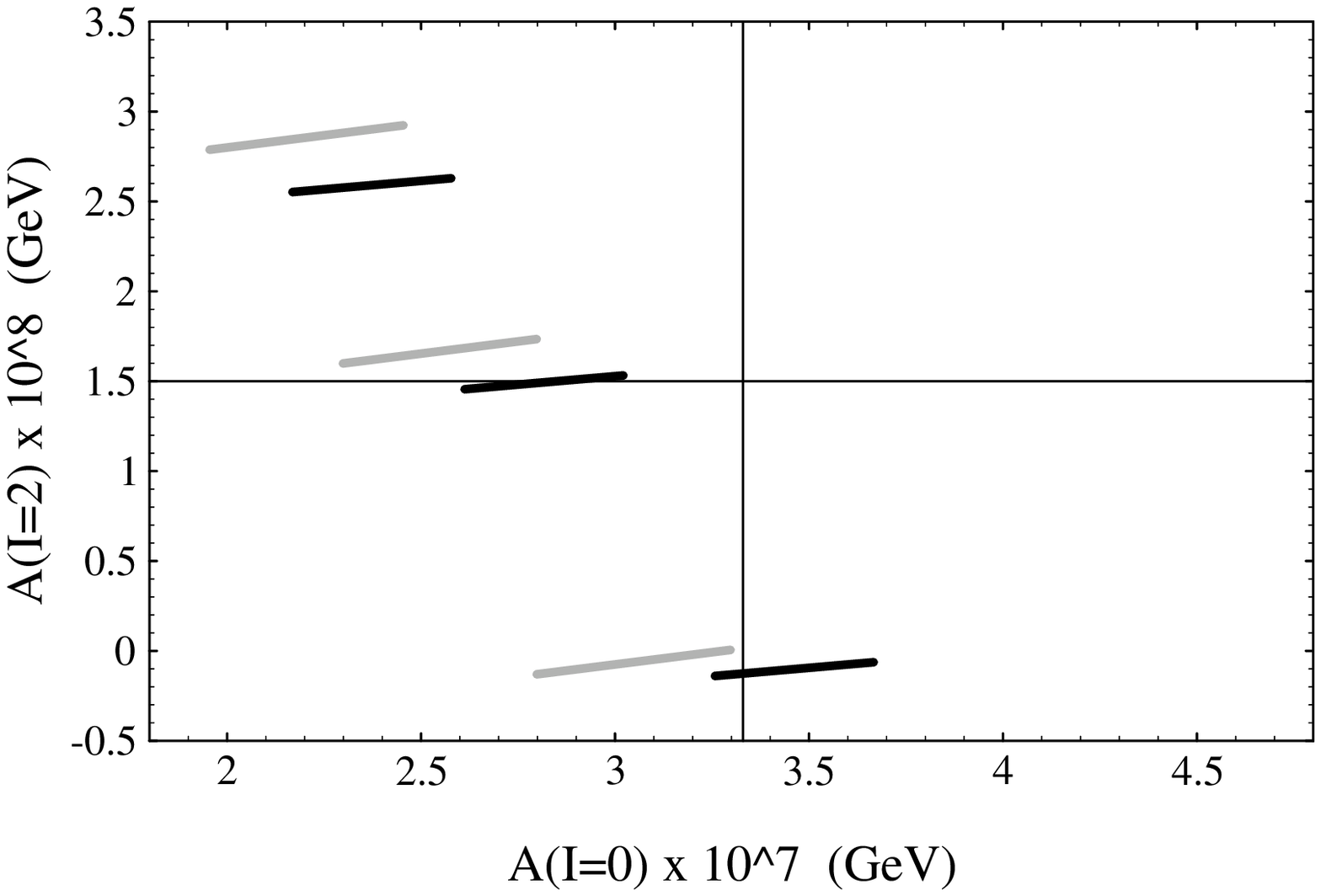}}
\caption{Same as in fig. 1 for $M = 220$ MeV.}
\end{figure}

The size of $A_2$ is a sensitive function of the gluon condensate and the
experimental value is approximately
reproduced in both schemes by the central value of
\eq{GGexp}.

Figs. 1, 2, 3 and 4 represent the main result
of this paper. For any choice of $M$ in the range 
\beq
M = 160 - 220 \quad \mbox{MeV}
\eeq
the experimental values of $A_0$ and $A_2$ are reproduced by our computation
by different choice of the input parameters $\vev{\bar{q}q}$ and $\vev{GG}$ 
that, however, remain within the
ranges discussed in section 3.1. Values of $M$ smaller than 160 MeV are 
still
consistent with the experiments, while the suppression of the gluon penguin 
contributions
as we increase $M$ above 220 MeV would force us to quark condensate
values outside the range of \eq{range}
in order to remain close to the experimental value 
for $A_0$.  

\clearpage
\subsection{$\gamma_5$-scheme independence}

The Wilson coefficients in \eq{ham} depend at the NLO on the
$\gamma_5$ scheme employed~\cite{Monaco,Roma}. In the $\chi$QM also
the hadronic matrix elements depend on the $\gamma_5$-scheme used in 
the computation. The requirement that these two dependences balance 
each other to provide a $\gamma_5$-scheme independent result 
 can be used in order to restrict the allowed values for the
free parameter $M$. 

The scheme dependence is not strong in  the amplitudes $A_0$ and
$A_2$, and, therefore, in this case, we can only restrict a range of values of
$M$ for which it is reasonably under control, but nothing more. 
The idea is
however that when examining other observables 
with a stronger scheme dependence as, 
for instance, 
$\varepsilon '/\varepsilon$, we may find a substantial
reduction of the scheme dependence 
for values
of $M$ that are still within the range 
determined by means of $A_0$ and $A_2$. A consistent picture for the whole
of kaon physics would thus emerge, with different observables concurring
in a consistent
determination of the free parameter $M$ as well as the range of the input
parameters. 

In our estimate of the hadronic matrix elements,
the $\gamma_5$-scheme independence of $A_0$ 
turns out to be controlled by the gluonic penguins. In fact,
the Wilson coefficients $z_{3-6}$ are systematically larger
in the NDR scheme than in the HV scheme. 
On the other hand, in the NDR scheme the matrix elements
of the gluonic penguins decrease with increasing $M$ faster than 
in the HV scheme.
As a consequence, it always exists
 a value of $M$ for which the $\gamma_5$-scheme independence
is achieved.

Larger values of the quark
condensate gives stability for larger values of $M$. From fig. 5 and 6 one
can readily see that the difference between the two schemes remains below 20\%
for  $M$ between 140 and 240 MeV. 
\begin{figure}[ht]
\epsfxsize=12cm
\centerline{\epsfbox{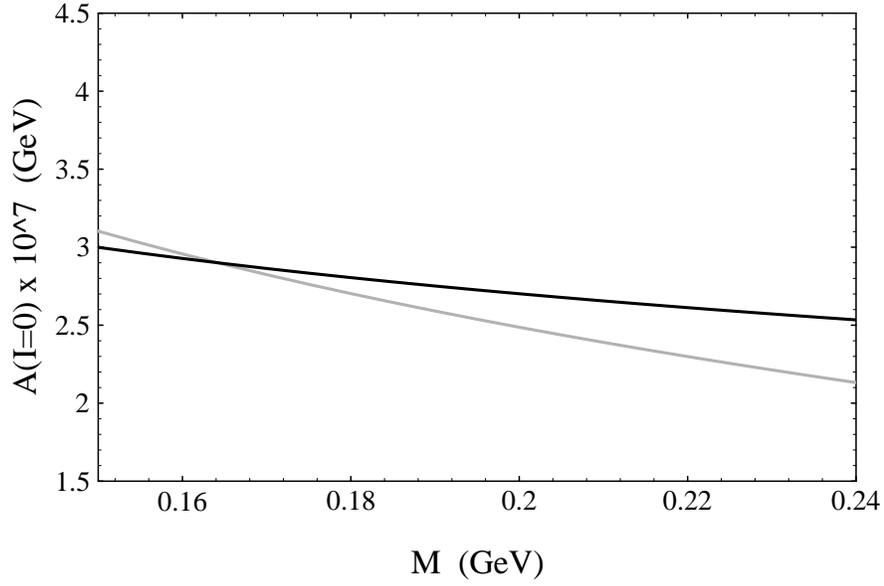}}
\caption{The $\gamma_5$-scheme dependence of $A_0$ is shown as
a function of $M$. The black (grey) line represent the HV (NDR)
result. 
We use $\vev{\alpha_s GG/\pi} = (376\ \mbox{MeV})^4$ and 
$\vev{\bar{q}q} = -(220\ \mbox{MeV})^3$. 
For this value of the quark condensate the stability 
appear near $M=160$ MeV. In the whole range of $M$ shown, the
$\gamma_5$-scheme dependence is always below 20\%. }
\end{figure}
\begin{figure}
\epsfxsize=12cm
\centerline{\epsfbox{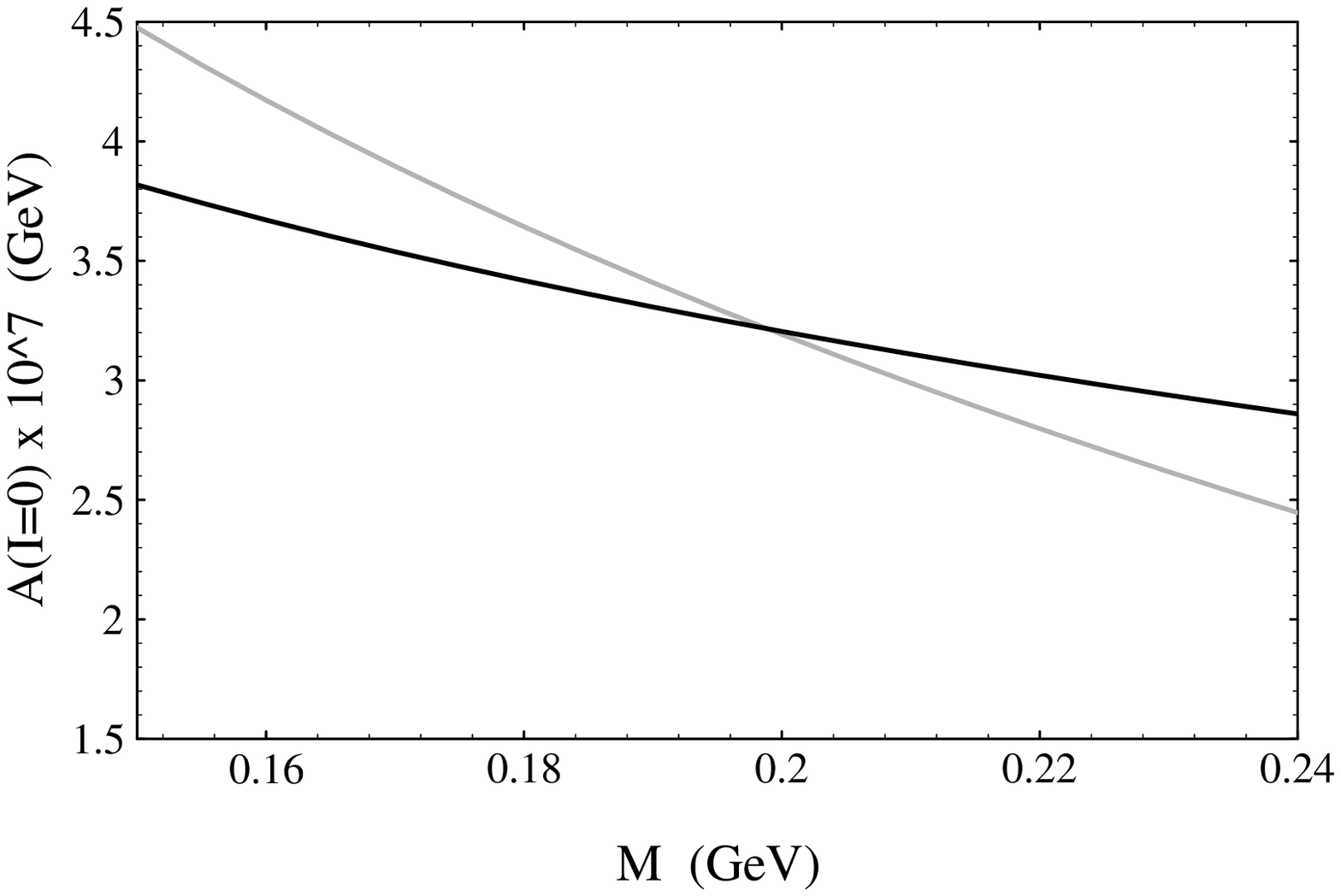}}
\caption{Same as in fig. 5 for $\vev{\bar{q}q} = -(280\ \mbox{MeV})^3$.  
The stability is moved at about $M=200$ MeV. }
\end{figure}

The same is not true for the amplitude $A_2$. In this case, the amplitude is
controlled by the operators $Q_1$ and $Q_2$ only so that there is
 no dependence on $M$ and, accordingly, no intersection between the
 two schemes. 
 The scheme dependence, however, stays well below 20\%, as it can be seen 
 in Fig. 7.
\begin{figure}
\epsfxsize=12cm
\centerline{\epsfbox{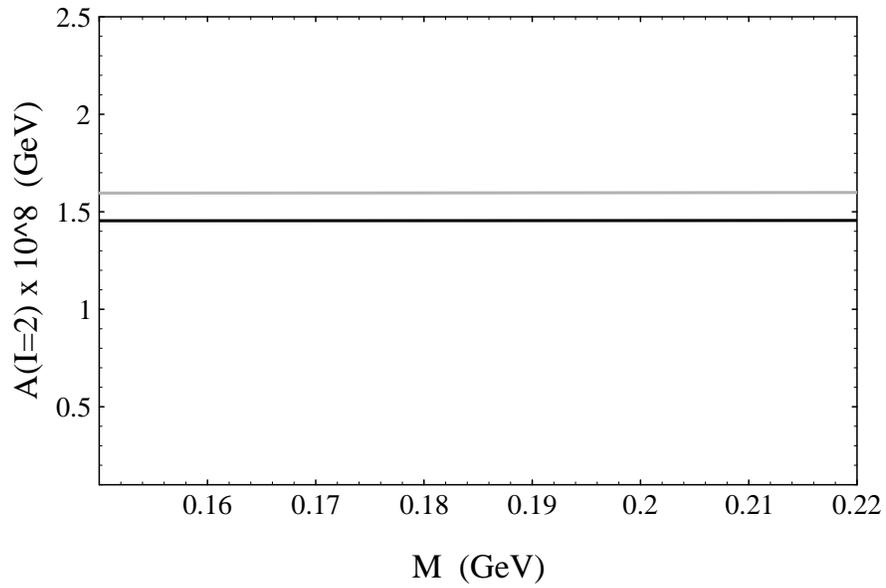}}
\caption{Same as in fig. 5 for $A_2$.}
\end{figure}

\clearpage
\section{The Road to the $\Delta I = 1/2$ Selection Rule}

An instructive way of analyzing the relevance of the various contributions
to the $\Delta I = 1/2$ rule is obtained by 
turning on in our computation each of them as
we follow the historical steps that have lead to the 
present understanding of the
rule (fig. 8).

The point labeled by (1) in fig. 8 represents the theoretical prediction
as obtained by considering the pure VSA matrix elements of $Q_1$ and $Q_2$
without the short distance renormalization of the corresponding Wilson
coefficients $(\mu=m_W)$. 
Point (2) represents the inclusion of
the NLO renormalized Wilson coefficients, matched to the 
hadronic matrix elements 
at the scale $\mu=0.8$ GeV. This scale
is large enough to make the renormalization-group analysis
reliable while keeping the hadronic matrix elements in the chiral
regime. 

As we can see, the value for $A_0$ is far too small and that of 
$A_2$ too large by a factor of two.

The introduction of penguin operators (point (3))
goes in the direction of increasing $A_0$, but it 
leaves $A_2$ unchanged. Their 
effect on $A_0$ is not, at least in the $\chi$QM, as crucial as often
claimed.

The introduction of the electroweak penguins ($Q_{7-10}$) little affects
the $CP$-conserving amplitudes (point (4)),
being suppressed
by the smallness of their Wilson coefficients. 
Another 
isospin breaking contribution to the amplitude $A_2$
comes from a long-distance effect, namely
the mixing between
$\pi^0$ and $\eta$ particles. This contribution is evaluated to be
\beq
A_2^{iso-brk} \simeq -\frac{1}{3 \sqrt{2}} \frac{m_d - m_u}{m_s} A_0 \, .
\label{isobrk}
\eeq
Accordingly, we have a reduction of the amplitude $A_2$
represented by point (5) which compensates for the effect
of the electroweak penguins. 
Because of the smallness of $A_0$ at point (5), the
effect of (\ref{isobrk})
is very small, though it reaches the 10\% level in the final point (7). 

A crucial
step toward the understanding of the selection
rule is due to the (non-factorizable) gluon-condensate corrections
(point (6)). 
They represent a genuine non-perturbative part of the computation.
The isospin asymmetry generated by the electroweak operators
is amplified in the right direction, improving
dramatically $A_2$, that becomes close to its experimental
value, while $A_0$ is further increased with respect to point (5).
The relevance of these contributions 
was first pointed out in ref. \cite{PdeR}.

\begin{figure}[ht]
\epsfxsize=12cm
\centerline{\epsfbox{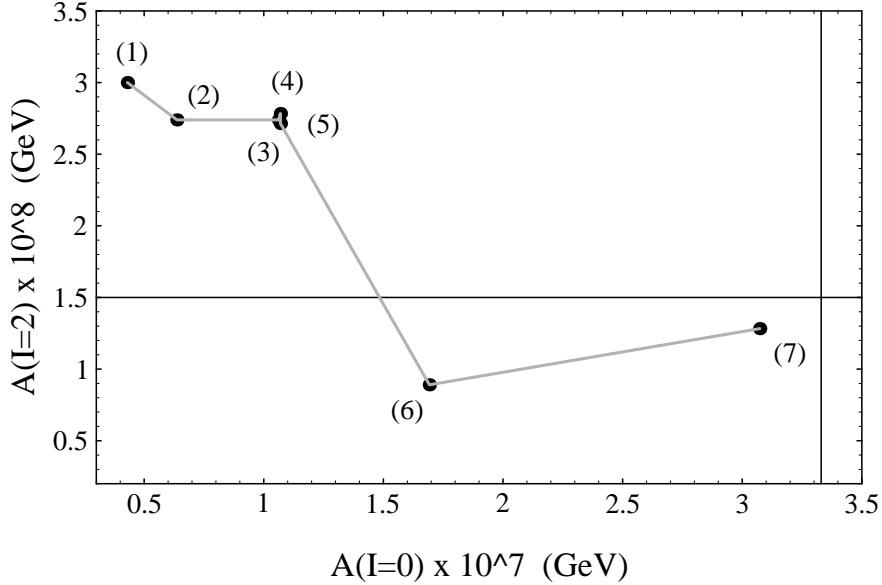}}
\caption{The road to the $\Delta I = 1/2$ rule:
(1) Effect of the W-induced current-current matrix elements 
   $(Q_{1,2})$ with the neglect of short-distance QCD renormalization
   $(\mu=m_W)$;
(2) $\vev{Q_{1,2}}$ with the inclusion 
   of the NLO Wilson coefficients at $\mu=0.8$ GeV.
(3) inclusion of the gluon penguins $(Q_{3-6})$;
(4) inclusion of the electro-weak penguins $(Q_{7-10})$;
(5) inclusion of the $\pi^0-\eta$ mixing;
(6) inclusion of gluon condensate corrections;
(7) meson-loop renormalization.
The results shown are those of the HV scheme with the central values 
$\vev{\alpha_s GG/ \pi} = (376 \: \mbox{MeV} )^4$ and
$\vev{\bar{q}q} = - (250\: \mbox{MeV} )^3$,  for a matching scale
$\mu = 0.8$ GeV and $M = 180$ MeV.
The experimental values are given by the cross hairs.}
\end{figure}

The meson-loop renormalization provides in our approach the final
step toward the experimental results.
The size of the relative renormalizations of 
$A_0$ and $A_2$ goes in the right direction,
being large for $A_0$ and small for
$A_2$. The meson loops also introduce a renormalization scale
dependence in the 
hadronic matrix elements to be matched with that of the
Wilson coefficients.

As point (7) of fig. 8 shows, the $\Delta I = 1/2$ selection rule is well
reproduced in the $\chi$QM, that provides not only values for the amplitudes
$A_0$ and $A_2$ that are close to the experimental ones but also a satisfactory
scale and $\gamma_5$-scheme independence of the estimate. 
Within 20\%
we also have scale independence in the 
matching range between 0.8 and 1 GeV.

In drawing fig. 8 we have taken 
the gluon and quark condensates at the 
central values of \eqs{GGexp}{qqexp}. 
Had we chosen, for instance, 
$\vev{\alpha_s GG/\pi} = (372 \: \mbox{MeV} )^4$ and
$\vev{\bar{q}q} = - (271\: \mbox{MeV} )^3$ we would have exactly reproduced
the experimental result. 

\bigskip \bigskip \bigskip

{\sc Acknowledgments}

\bigskip
We thank our collaborator J.O. Eeg for
teaching us about the $\chi$QM and  for many discussions. 

This work was partially supported by the EEC Human Capital and
Mobility  contract ERBCHRX CT 930132.

\newpage

\appendix
\section{Input Parameters} 
\begin{table}[h]
\begin{center}
\begin{tabular}{|c|c|}
\hline
{\rm parameter} & {\rm value} \\
\hline
$V_{ud}$ & 0.9753 \\
$V_{us}$ & 0.221 \\
$\sin ^2 \theta_W$ & 0.2247 \\
$m_Z$ & 91.187 GeV \\
$m_W$ & 80.22 GeV \\
$m_b$ & 4.8 GeV \\
$m_c$ & 1.4 GeV \\
\hline
$f_\pi = f_{\pi^+}$  &  92.4  MeV \\
$f_K = f_{K^+}$ & 113 MeV \\
$m_\pi = (m_{\pi^+} + m_{\pi^0})/2 $ & 138 MeV \\
$m_K = m_{K^0}$ &  498 MeV \\
$m_\eta$ & 548 MeV \\
$\Lambda_\chi$ & $2 \sqrt{2} \pi f_\pi$ \\
\hline
$\Lambda_{QCD}^{(4)}$ & $350 \pm 100$ MeV \\
$\overline{m}_u + \overline{m}_d$ (1 GeV) & $12 \pm 2.5$ MeV \\
$\vev{\bar{q}q}$  &  $- (200 -280 \: \mbox{MeV} )^3$ \\
$ \langle \alpha_s GG/\pi \rangle $ & $(376 \pm  47 \: 
\mbox{MeV} )^4 $ \\
\hline
\end{tabular}
\end{center}
\caption{Table of the numerical values of the input parameters.}
\end{table}
%
%
\clearpage
\renewcommand{\baselinestretch}{1}

\end{document}